\documentclass[twocolumn,showpacs,amsmath,superscriptaddress,amssymb,prb,floatfix,10.75pt]{revtex4-2}
\usepackage{graphicx}  
\usepackage{epsfig}    
\usepackage{dcolumn}   
\usepackage{bm}        
\pdfoutput=1 
\usepackage{amssymb}   
\usepackage{amsmath}   
\usepackage{mathtools} 
\usepackage{natbib}    
\usepackage{lipsum}
\setcitestyle{super}
\usepackage{xcolor}
\usepackage{array}
\newcolumntype{P}[1]{>{\centering\arraybackslash}p{#1}}
\usepackage{longtable}
\usepackage{tabularray}

\usepackage[hidelinks]{hyperref}
\usepackage{longtable,booktabs}
\usepackage{marvosym}

\usepackage{wrapfig}
\usepackage[acronym]{glossaries}
\makeglossaries
\newacronym{dt}{DT}{decision tree}
\newacronym{rf}{RF}{random forest}
\newacronym{gb}{GB}{gradient boosting}
\newacronym{xgb}{XGB}{extreme gradient boosting}
\newacronym{mse}{MSE}{mean squared error}
\newacronym{ann}{ANN}{artificial neural network}
\newacronym{dft}{DFT}{density functional theory}
\newacronym{ml}{ML}{machine learning}
\newacronym{ct}{$T_{\rm C}$}{Curie temperature}
\newacronym{hc}{$H_c$}{coercivity}
\newacronym{ms}{$M_s$}{saturation magnetization}
\newacronym{ku}{$K_u$}{uniaxial magnetocrystalline anisotropy constant}
\newacronym{aex}{$A_{\mathrm{ex}}$}{exchange stiffness constant}
\newacronym{lgbm}{LGBM}{light gradient boosted machine}
\newacronym{mae}{MAE}{mean absolute error}
\newacronym{vif}{VIF}{variance inflation factor}
\newacronym{rmse}{RMSE}{root mean square error}
\newacronym{soc}{SOC}{spin-orbit coupling}
\newacronym{pbe}{PBE}{Perdew-Burke-Ernzerhof}
\newacronym{vasp}{VASP}{Vienna simulation package}
\newacronym{paw}{PAW}{projector augmented wave}
\newacronym{gga}{GGA}{generalized gradient approximation}
\newacronym{asa}{ASA}{atomic sphere approximation}
\newacronym{lda}{LDA}{local density approximation}
\newacronym{lmto}{LMTO}{linearized muffin-tin orbital}
\newacronym{mfa}{MFA}{mean-field approximation}
\newacronym{rpa}{RPA}{random-phase approximation}
\newacronym{llg}{LLG}{Landau–Lifshitz–Gilbert}
\newacronym{dmi}{DMI}{Dzyaloshinskii–Moriya interaction}

\usepackage[utf8]{inputenc}
\usepackage{float}
\usepackage{newunicodechar}
\newunicodechar{−}{-}
\usepackage{rotating}

\newcommand{\TheorySize}{$8770$}
\newcommand{\ExpSize}{$300$}
\newcommand{\mumax}{$\mathrm{mumax}^3$~}

\begin{document}

\title{Accurate Machine Learning Predictions of Coercivity in High-Performance Permanent Magnets}

\author{Churna Bhandari}
\email{cbb@ameslab.gov}
\affiliation{The Ames National Laboratory, U.S. Department
of Energy, Iowa State University, Ames, Iowa 50011, USA}
\author{Gavin N. Nop}
\affiliation{The Ames National Laboratory, U.S. Department
of Energy, Iowa State University, Ames, Iowa 50011, USA}
\affiliation{Department of Mathematics,  Iowa State University, Ames, Iowa 50011, USA}
\author{Jonathan D.H. Smith}
\affiliation{The Ames National Laboratory, U.S. Department
of Energy, Iowa State University, Ames, Iowa 50011, USA}
\affiliation{Department of Mathematics,  Iowa State University, Ames, Iowa 50011, USA}
\author{Durga Paudyal}
\affiliation{The Ames National Laboratory, U.S. Department of Energy, Iowa State University, Ames, Iowa 50011, USA}
\affiliation{Department of Electrical and Computer Engineering,  Iowa State University, Ames, Iowa 50011, USA}

\date{December 24, 2023}


\begin{abstract}
Increased demand for high-performance permanent magnets in the electric vehicle and wind turbine industries has prompted the search for cost-effective alternatives. Discovering new magnetic materials with the desired intrinsic and extrinsic permanent magnet properties presents a significant challenge to researchers because of issues with the global supply of rare-earth elements, material stability, and a low maximum magnetic energy product $BH_{max}$. While  first-principle \gls{dft} predicts materials' magnetic moments, magneto-crystalline anisotropy constants, and exchange interactions, it cannot compute extrinsic properties such as \gls{hc}. Although it is possible to calculate $H_c$ theoretically with micromagnetic simulations, the predicted value is larger than the experiment by almost an order of magnitude, due to the Brown paradox. To circumvent these issues, we employ \gls{ml} methods on an extensive database obtained from experiments, \gls{dft} calculations, and micromagnetic modeling. The use of a large experimental dataset enables realistic $H_c$ predictions for materials such as $\mathrm{Ce}$-doped $\mathrm{Nd}_2\mathrm{Fe}_{14}\mathrm{B}$, comparing favorably against micromagnetically simulated coercivities. Remarkably, our \gls{ml} model accurately identifies uniaxial magneto-crystalline anisotropy as the primary contributor to $H_c$. With \gls{dft} calculations, we predict the Nd-site dependent magnetic anisotropy behavior in $\mathrm{Nd}_2\mathrm{Fe}_{14}\mathrm{B}$, confirming that $\mathrm{Nd}$ $4g$-sites mainly contribute to uniaxial magneto-crystalline anisotropy, and also calculate the \gls{ct}. Both calculated results are in good agreement with experiment. The coupled experimental dataset and \gls{ml} modeling with \gls{dft} input predict \gls{hc} with far greater accuracy and speed than was previously possible using micromagnetic modeling. Further, we reverse-engineer the grain-boundary and inter-grain exchange coupling with micromagnetic simulations by employing the ML predictions.
\end{abstract}

\maketitle

\section{Introduction}
With the rapid advance of computational capabilities, there is considerable research interest in machine learning (\gls{ml}) methods for predicting material properties using extensive databases\cite{MLbook, MLbook1}. Of specific interest is the remarkable speed of these techniques, which outperform traditional first-principle methods like density functional theory (\gls{dft})  by an order of magnitude. \gls{ml} methods can deal with complex structures, and are desirable for discovering high-performance permanent magnet materials much needed for the electric vehicle and wind turbine industries.  Although recent advances in first-principle methods such as \gls{dft} have enabled successful prediction of intrinsic properties, e.g., magnetic moments, magneto-crystalline anisotropy, and exchange interactions, the prediction of coercivity (\gls{hc}) is a daunting task. Theoretically, \gls{hc} can be computed by solving the phenomenological Landau–Lifshitz–Gilbert equation (LLGE) with micromagnetic simulations, but these simulations overestimate the experimental \gls{hc} by an order of magnitude due to the Brown paradox\cite{BrownparadoxRevModPhys45, AharoniRevModPhys62, HartmannPRB87}. Previous papers employing \gls{ml} to predict the extrinsic properties\cite{XiePRL018, FaberPRL016, Xue2016, SekoPRL015}  have been limited  exclusively to micromagnetically simulated materials. 

\gls{ml} requires datasets that include information about material  properties such as crystal structure, micromagnetic grain size and boundaries, \gls{ms}, the \gls{ku}, the \gls{aex}, and the Curie temperature (\gls{ct}). To build predictive models based on our dataset, we utilize classical \gls{ml} and \gls{ann} algorithms\cite{HopfielPNAS82,Hertz91,FreanNC90}. These models establish patterns and relationships between the independent and dependent (\gls{hc}) material properties. They are trained on subsets of the known data and tested on the complementary subsets to assess model accuracy and reliability. 

An extensive survey of the  literature resulted in a dataset of \ExpSize~experimentally known materials (see Supplementary\cite{supplemtary} Table I), to our knowledge the largest current experimental \gls{ml} magnetic dataset. Our second dataset consists of \TheorySize~micromagnetically computed permanent magnet materials. Various predictive techniques, including \gls{ml}, statistical inference, and micromagnetic modeling (\texttt{mumax$^3$} program) are applied to both datasets to predict and compare \gls{hc} \cite{Liu2012,SCHMIDHUBER201585}. In experimental materials, we find standard non-linear models such as the  \gls{dt}, \gls{xgb}, and \gls{rf}\cite{ho1995random} produce excellent results with $R^2 \sim 0.87$ (where $R^2$ is a standard statistical measure of accuracy in regression), but tuning the \gls{xgb} regressor improves the $R^2$ measure to $0.89$. Most importantly, \gls{ml} clearly demonstrates that \gls{hc} is related directly to \gls{ku}, weakly to \gls{aex}, and inversely to \gls{ms}.

We predict the \gls{hc} of cerium (Ce)-doped Nd$_2$Fe$_{14}$B 2\,:\,14\,:\,1 materials to demonstrate the complete pipeline enabled by the new \gls{ml} toolkit  coupled with \textit{ab initio} calculations. First, for a pure neo-magnet, the site contribution to magnetocrystalline anisotropy is analyzed with \gls{dft} calculations showing that $4g$-sites mainly contribute to the uniaxial magneto-crystalline anisotropy. Second, our computed $T_{\rm C}$ for a pure compound using Green's function in the \gls{asa} is in good agreement with experiment. Finally, we employ DFT-computed parameters in ML for predicting the $H_c$ of Ce-doped compositions. The \gls{ml}-predicted $H_c$ matches with experiment, demonstrating that the \textit{ab initio} computed input parameters and the \gls{ml} methodology are sufficient to predict experimental \gls{hc}, even without access to experimental conditions and advanced internal structural properties. Conversely, the ML prediction for selected candidate materials is used to engineer their grain boundary size (GBS) and inter-grain coupling.

\section{Micromagnetism}

Micromagnetics is the study of the behavior of magnetic materials typically in the nanometer range. During its early formulation\cite{GilbertIEEE04, Brown1963}, the field emphasized qualitative aspects of magnetism:  the role of domain structures, domain walls, and magnetic vortices in ferromagnetic materials. The transition to computer simulation in micromagnetics 
was a significant advance, providing detailed examination of the forces at play inside a magnetic material \cite{kronmuller2006general}.

 Naively, the estimate $\dfrac{2K_u}{M_s}$ gives an upper bound on \gls{hc} \cite{BrownparadoxRevModPhys45, HerbstRevModPhys91}. However, this estimate disregards impurities and multi-grain structures in materials, leading to a gross overestimate of \gls{hc} from theory alone: the Brown paradox. Micromagnetic simulations which include demagnetization (shape anisotropy) are crucial for a good understanding of magnetic materials, but the interplay among these complex effects and the Brown paradox still hinder accurate predictions of \gls{hc}. 

\subsection{Theory}

Magnetodynamics is described by a nonlinear partial differential equation for the spatio-temporal magnetization vector ${\bf M}({\bf r},t)$.
The time evolution of ${\bf M}({\bf r},t)$ is given by a phenomenological Landau–Lifshitz–Gilbert (LLG) equation \cite{Brown1963,GilbertIEEE04}
\begin{eqnarray}
\dfrac{\partial {\bf M}({\bf r},t)}{\partial t}& = &\dfrac{\gamma}{1+\alpha^2}{\bf M}({\bf r},t)\times {\bf H_{\mathrm{eff}}} ({\bf r},t)\nonumber \\ 
& & - \dfrac{\alpha\gamma}{1+\alpha^2} {\bf M}({\bf r},t) \times\big[{\bf M}({\bf r},t)\times {\bf H_{\mathrm{eff}}}({\bf r},t)\big].
\label{GL}
\end{eqnarray}
Here, ${\bf M}({\bf r},t)$ is the unit vector describing the magnetization of the material with \gls{ms} as the saturation magnetic moment per unit volume, while ${\bf H_{\mathrm{eff}}}({\bf r}, t)$, $\gamma$, and $\alpha$ respectively are the effective static magnetic field, the gyromagnetic ratio, and the damping parameter (quantifying the rate at which the magnetization relaxes back to equilibrium). In Eq.~\eqref{GL}, the first term is the precession of the magnetic moment around the external magnetic field. The second term is the damping, which relaxes the magnetic moment to the equilibrium. For time-independent scenarios, such as the computation of a hysteresis loop, the $\alpha$ term is set to $0$. Then, ${\bf H_{\mathrm{eff}} = H + H_{\mathrm{ms}} + H_{\mathrm{ex}}+ H_{a}}$. The terms are as follows. ${\bf H}$ is the externally applied field, which is taken as a parameter. ${\bf H_{\mathrm{ms}}}$ is a long-range magnetic field
\begin{equation}
    {\bf H_{ms}({\bf r})}= \dfrac{1}{4\pi}\int \nabla \nabla^{'}\dfrac{1}{{\bf |r-r^{'}|}}. M({\bf r^{'}})dr^{'}
\end{equation}
corresponding to self-interaction of the induced magnetic field with the magnetization across the material \cite{aharoni2000introduction}. The exchange field is the Laplacian of the magnetization, which is obtained from the classical Heisenberg model
\begin{equation}
        {\bf H_{\mathrm{ex}}} = \dfrac{2A_{\mathrm{ex}}}{M_s}\nabla {\bf M}({\bf r},t).
\end{equation}
Although this expression was originally deduced for localized spins, it is still valid for itinerant systems to the first order approximation. $A_{\mathrm{ex}}$ is a measure of the strength of magnetic exchange interaction. $H_{a}$ is the uniaxial anisotropy term, written as 
\begin{equation}
    {\bf H_{a}} = \frac{2K_{u1}}{M_{s}}\big({\bf u} \cdot {\bf M}({\bf r},t)\big){\bf u} + \frac{2K_{u2}}{M_{s}}\big({\bf u} \cdot {\bf M}({\bf r},t)\big)^3{\bf u}
\end{equation}
where $K_{u1}$ and $K_{u2}$ are anisotropy constants, and $u$ indicates the direction of the anisotropy vector, making it easier for the magnetization to align with the direction of $u$ in the case of ferromagnetic materials. Most papers only envisage a single constant: $K_{u1} = K_u$ and $K_{u2} = 0$. Other terms, such as the Dzyaloshinskii-Moriya interaction (DMI), were neglected due to the lack of experimental data for current materials and the relatively weak effect in modern magnetic materials \cite{laplane2016high}. 

In order to estimate fundamental magnetic parameters in cases of partial experimental information, the following relations were utilized. $A_{\mathrm{ex}}$ is inversely  proportional to the lattice constant ($a$) as given by
\begin{eqnarray}
    A_{ex}= \dfrac{JS^2}{a}n,
    \label{aex1}
\end{eqnarray}
where $S$ is the spin quantum number, and $n$ is the number of magnetic ions per unit cell. Equivalently, $A_{\mathrm{ex}}$ can be expressed approximately in terms of $T_{\rm C}$ as
\begin{eqnarray}
    A_{ex}\sim \dfrac{3T_{\rm C}}{2za}.
    \label{aex2}
\end{eqnarray}
Here $z$ is the number of the nearest neighbors of a magnetic ion. Finally, the equation 
\begin{eqnarray}
    BH_{\mathrm{max}} = \frac{\mu_0 M_s^2}{4}
    \label{aex3}
\end{eqnarray}
is used to estimate  \gls{ms} from the vacuum permeability $\mu_0$ and the maximum energy product $BH_{\mathrm{max}}$, which is given by the maximum product of the magnetic flux density and the magnetic field strength at any point on the hysteresis loop \cite{bhandari2023giant}.

The micromagnetic equations are  solved in the continuum approximation ${\bf M}({\bf r},t)=M_s(r) {\bf m}({\bf r},t)$\cite{Vansteenkiste2014, Lel2014, Exl2014}.
${\bf H_{eff}}$ is deduced from the magnetic free energy functional $F[{\bf m}]$ as ${\bf H_{eff}} =-\dfrac{1}{\mu_0 M_s} \dfrac{\delta F[{\bf m}]}{\delta {\bf m}}$;

\begin{eqnarray}
    F[{\bf m}] &= &\int_V \Big[ A_{ex}(\nabla {\bf m})^2 -\mu_0 {\bf M.H_{ex}} -K_u({\bf m.u})^2\nonumber \\ &&- \mu_0 {\bf M. H}- \dfrac{\mu_0}{2} {\bf M.H_{dem}} + f_{\rm{DMI}}({\bf m}) + \cdots \Big] d^3{\bf r},\nonumber\\
    \label{freeenergy}
\end{eqnarray}
where   
$f_{DMI}$ is DMI. Additional terms may be added as needed to count for additional interactions. The $F[m]$ is minimized with respect to $m$ using the steepest descent algorithm as implemented in the micomagnetic program \cite{Exl2014}.
Then the hysteresis loop is obtained by evaluating $m$ in each equilibrium magnetic state  for different values of the external applied magnetic field.

\begin{figure}[H]
\centering
\includegraphics[scale=0.2]{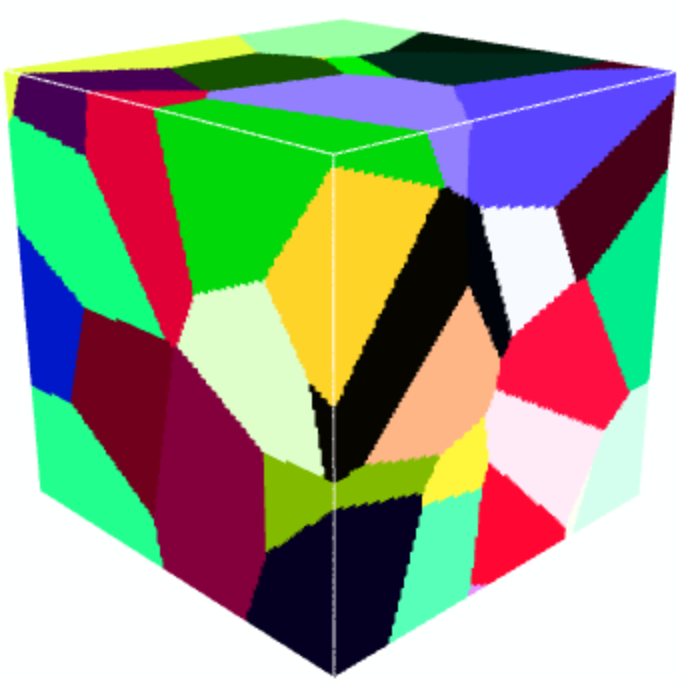}
\caption{A magnetic cuboid of size $128 \times 128 \times 128$ nm$^3$ showing different micromagnetic grains. The colors within each grain refer to different magnetization directions. The snapshot of spin texture is captured during the micromagnetic simulation.}\label{structure}
\end{figure}

\subsection{Experimental materials}
\label{databses}

An experimental database of \ExpSize ~magnetic materials was assembled\cite{Imamura68,Ristau99,HerbstRevModPhys91,Sales2014, JABALLAH2022110438,refId0,Liu21, SATO2022168684,YU2018272, KIKUCHI20102381,REHMAN201983,WANG2018193,ZHANG20123272, GABAY201475,HIRAYAMA201627,fliegans:tel-02635105,CHENG1998310, Oyama96, WU2023167888, Tsui03, ZHOU2014127,dGkouzia23, SURESH2007358,TELLEZBLANCO19981, Kuzmin015,Huang00,ONYSZCZAK2019192,WONG2020167287,SILVA2015263,KATLAKUNTA201558, CHAWLA201584,ALANGE2016460, Lee2020, RehmanACS022,Leessrn,Ghimire20,Aledealat21, Luo16, Azis2018, HeAVM022, STRNAT199138,KOVACS2020148, Juha, KOVACS2020148, NIEVES2019188, LI2015325,SUSNER20171,KOBAYASHI2014569,Shao-ying,FAN2016394,ZHOU20231058, LEE1999137,Fischbacher17,Liu23,Goll014,Bick13,Ohtani77}. A material was chosen if \gls{ku}, \gls{ms}, and \gls{hc} were explicitly included in the experimental results. 
For the R$_2$Fe$_{14}$B family (where R is a rare-earth element) and some other materials, only the experimental anisotropy field ($H_a$) is available, so the estimate $H_c \sim H_a/4$ was used\cite{osti4769190}.

The exchange strength \gls{aex} was determined in one of three different ways. The first was a direct reporting of the stiffness exchange, which was rarely available. For some materials, the value was interpolated from the values for similar alloys and compounds. The third method derived \gls{aex} from the Curie-stiffness relation in Eq.~\eqref{aex2}, using the experimental Curie temperature $T_{\rm C}$ and known lattice constant $a$\cite{InabaJMM01}. This latter method was used for R$_2$Fe$_{14}$B, including Ce-doped Nd$_2$Fe$_{14}$B\cite{SUSNER20171}, binary alloys\cite{NIEVES2019188}, SmCo$_{5}$, 1\,:\,5 compositions\cite{STRNAT199138}, La/Pr/Co-doped hexaferrites\cite{Ghimire20}, and other element doped hexaferrites\cite{ALANGE2016460,TayssirJAC016,CHAWLA201584, Leessrn}.    

There was significant variation in the reported  $H_c$ values, depending on the experimental conditions and fabrication methods. For example, the atmospheric composition during the annealing of iron (Fe) magnets affected $H_c$ as oxygen modified the material composition during the hardening process. 
Oxygenated crystalline defects, such as nucleation and pinning of the domain walls, play critical roles in the $H_c$ mechanism. The nucleation field\cite{nucleation} (the magnetic field at which the atomic spin ceases to align along the magnetic easy axis) lowers the measured $H_c$, while pinning does the opposite.

Cooling rates have significant effects on the material grain size\cite{Imamura68,Ristau99,HerbstRevModPhys91,Sales2014, JABALLAH2022110438,refId0,Liu21, SATO2022168684,YU2018272, KIKUCHI20102381,REHMAN201983,WANG2018193,ZHANG20123272, GABAY201475,HIRAYAMA201627,fliegans:tel-02635105,CHENG1998310, Oyama96, WU2023167888, Tsui03, ZHOU2014127,dGkouzia23, SURESH2007358,TELLEZBLANCO19981, Kuzmin015,Huang00,ONYSZCZAK2019192,WONG2020167287,SILVA2015263,KATLAKUNTA201558, CHAWLA201584,ALANGE2016460, Lee2020, RehmanACS022,Leessrn,Ghimire20,Aledealat21, Luo16, Azis2018, HeAVM022, STRNAT199138,KOVACS2020148, Juha, KOVACS2020148, NIEVES2019188, LI2015325,SUSNER20171,KOBAYASHI2014569,Shao-ying,FAN2016394,ZHOU20231058, LEE1999137,Fischbacher17,Liu23,Goll014,Bick13,Ohtani77}. Several dozen materials (most notably Nd and Fe based ferromagnets) do have grain size recorded in the database. Grain composition, as visualized in Fig.~\ref{structure}, is known to be heavily predictive of a material's $H_c$\cite{GrainHerzer90}. However, correlating the grain measurements from different sources made it apparent that a single number cannot completely represent the grain structure, and that the particular measurement techniques employed may induce additional inaccuracies. In any material with multiple measurements for a property, a representative median was selected from a specific experiment (defaulting to papers with more comprehensive materials property estimates). This choice was made before the application of statistical techniques, to ensure that no bias contaminated the dataset.

For the better training of \gls{ml} networks, an additional supplementary database was computed, using micromagnetic modeling with $H_{\mathrm{ms}}$, $H_{\mathrm{ex}}$, $H$, $H_{\mathrm{dem}}$, and $H_{a}$ in the $H_{\mathrm{eff}}$. Using the uniaxial approximation,  $A_{ex}$, $K_u$, and $M_s$ were supplied as the inputs for stiffness exchange, magnetocrystalline anisotropy, and saturation magnetization. The input parameters were uniformly sampled from the cuboid determined by the ranges of $A_{ex}$, $K_u$, and $M_s$ in the \ExpSize~experimental materials. 
There are then two measures of $H_c$: the experimentally measured coercivity $H_c$(exp), possessed only by the experimental materials, and the computationally determined $H_c$, calculated for both the hypothetical and the experimental materials.

For micromagnetic modeling, simple structures were chosen for the magnetic samples. Due to the limited grain structure data, for the majority of the database a single structure composed of multiple grains would have been chosen for the materials. This would have induced scaling on the \gls{hc}, which instead may be modeled directly with \gls{ml}. As no grains or boundaries were involved in the computation, a magnetic cube of $32\times 32\times 32$ at nm$^3$ scale was found to be sufficient to avoid loss of precision in the uniform uniaxial anisotropy alignment. For Ce-substituted compositions, we used simulation cells of size $128\times 128\times 128$ nm$^3$ for grain boundary engineering as shown in Fig.~\ref{structure}.


\subsection{Micromagnetic results}

We compute the $H_c$ for experimentally known systems, and list the comparison of the calculated and experimental values for selected materials in Table~\ref{tabcoercivity}. Theoretically, the upper limit for $H_c$ is the anisotropy field\cite{SKOMSKI20163} $H_a=\dfrac{2K_u}{M_s}$; however, the true experimental value is an order of magnitude smaller due to the uncertainty in the coercivity mechanism in permanent magnets, commonly known as the Brown paradox \cite{BrownparadoxRevModPhys45}. Experimentally measured $H_c$ values fit well
with the empirical Kronm\"uller equation\cite{Kronmpss1987} $H_c= c H_a-N_{\mathrm{eff}} 4\pi M_s$\cite{PinkertonJAP90, HERBST199157}, where $c H_a$ is the field to nucleate a reverse domain, and $N_{\mathrm{eff}} 4\pi M_s$ is the demagnetization field, with $c$ and $N_{\mathrm{eff}}$ being renormalization factors. For example, Nd$_2$Fe$_{14}$B (sintered) has $c\sim 0.25(0.37)$  and $N_{\mathrm{eff}}\sim 0.26(1)$. \cite{PinkertonJAP90, HERBST199157}

\begin{table*}
\centering
\begin{ruledtabular}
\caption{Comparison of calculated and experimental \gls{hc} for different rare-earth permanent magnets. For realistic comparisons, the \mumax values are scaled by a factor of $c=0.25$ in column $cH_c$. The $H^{\mathrm{diff}}_c$ column presents the difference between $cH_c$ and $H^{\mathrm{exp}}_c$.}
\begin{tabular}{l | l c l l l l l l}
Material & $M^{\mathrm{exp}}_s$ (MA/m) & $A^{\mathrm{exp}}_{ex}$ (pJ/m) & $K^{\mathrm{exp}}_u$ (MJ/m$^3$) & $H^{\mathrm{exp}}_c$ (T) & $H^{\mathrm{mumax}^3}_c$ (T) & $cH_c$ (T)&$H^{\mathrm{diff}}_c$ (T)\\
\hline

La$_2$Fe$_{14}$B	& 1.09	&7.41	&1.4	&0.50	&2.11	&0.53 & ~0.03\\
Ce$_2$Fe$_{14}$B	& 0.93	&6.00	&1.70	&0.54	&3.14	&0.79 &~0.25\\
Pr$_2$Fe$_{14}$B	& 1.24	&7.94	&4.66	&1.54	&6.46	&1.62 &~0.08\\
Nd$_2$Fe$_{14}$B	& 1.27	&8.23	&4.65 &1.24	&6.25	&1.56 &~0.32\\
Gd$_2$Fe$_{14}$B	&0.71	&9.33	&0.85	    &0.23	&2.40	& 0.60 &~0.37\\
Tb$_2$Fe$_{14}$B	&0.56	&8.78	&6.13	&2.93	&21.62	&5.41    &~2.48\\
Dy$_2$Fe$_{14}$B	&0.56	    &9.42	&4.24	&1.70	&14.79	&3.70 &~1.99\\
Ho$_2$Fe$_{14}$B	&0.64	&8.14	
&2.42	& 0.75	&7.26	& 1.82&~1.07\\

Lu$_2$Fe$_{14}$B	&0.93	&7.66	&1.21	&0.29 &2.30&0.58 &~0.28\\

Y$_2$Fe$_{14}$B	    &1.12	&8.01	&1.46	&0.19	&2.16	&0.54 &~0.35\\

Th$_2$Fe$_{14}$B	&1.12	&6.77	&1.46	& 0.17	&	2.10	&0.53 &~0.36\\

La$_2$Co$_{14}$B	&0.79	&1.50 	&1.19 &0.34	&	3.06	&0.77& ~0.42\\

Pr$_2$Co$_{14}$B	&1.04	&1.51 	&5.20	&2.50&	9.43 &2.36&-0.14\\

Nd$_2$Co$_{14}$B	&1.08	&1.51 	&2.42 &3.69	&4.22 &1.05 &-2.64\\

Gd$_2$Co$_{14}$B	&0.23	&1.52 	&1.03 &0.30	&	9.48&2.37& ~2.07\\

Y$_2$Co$_{14}$B	    &0.85	&1.52 	&1.19 &0.34	&	2.84	&0.71& ~0.37\\

SmCo$_5$ &	0.86	&1.20 	&1.72	&7.50	&38.80& 9.70 & ~2.20\\

YCo$_5$	&0.78	&7.11	&5.50	&3.90&	13.33	&3.33&-0.57\\

LaCo$_5$	&0.71	&6.47	&6.30	&5.25	&	16.89	&4.22 &-1.03\\

CeCo$_5$	&0.60	&5.13	&6.40	&5.70	&20.66	&5.17	&~0.57\\

PrCo$_5$	&0.94	&6.96	&8.10	&5.32	&	16.08&4.02	&-1.30\\

NdCo$_5$	&0.93	&7.09	&0.24	&0.15	&0.47	&0.12	&-0.03\\
\end{tabular}
\label{tabcoercivity}
\end{ruledtabular}
\end{table*}

The micromagnetically simulated \gls{hc} values differ from the experimental values by a factor of up to $\sim 5$. Interestingly, these values are very similar to the experimental $H_a$. Moreover, for some materials, the use of the estimated values for $M_s$, $A_{ex}$, and $K_u$ using the empirical relations as discussed in Eqs.~\eqref{aex1}--\eqref{aex3} will result in additional error. In general, \gls{hc} depends non-linearly on grain size and domain wall width or particle size in magnetic materials\cite{AlbenJAP08, AkdoganAFM13, HERZERActa095, HerzerIEE90, FischerJMM096}. In modern manufacturing, grain sizes are larger relative to domain sizes. This is especially common in neo-magnets, leading to a decrease of \gls{hc} with increasing grain size\cite{RameshJAP088}, where higher grain surface area hosts more defects. Grain boundary size also affects the demagnetization factor\cite{BanceJAP014}, further reducing \gls{hc}.

The experimental features for 211 of the \ExpSize~materials in the dataset are known, while the remaining 90 materials require the use of the aforementioned theoretical models to determine \gls{aex} from experimental results. These materials show larger discrepancies in the micromagnetically predicted \gls{hc}, which is worth investigating both experimentally and theoretically. We show a comparison of these results (for selected key magnetic materials) in Table~\ref{tabcoercivity}. As all micromagnetically predicted \gls{hc} values are overestimated, the only meaningful comparison is obtained with a scaled coercivity $cH_c$, which for $c=0.25$ fits well with the experimental values obtained for the 210 materials. The $cH_c$ estimate is generally a good match for 2\,:\,14\,:\,1 compositions, except for Co-based Nd$_2$Co$_{14}$B and Gd$_2$Co$_{14}$B. However, the appropriate scaling factor may vary for different compositions. Overall, the \gls{hc} variation with independent features is similar in both theory and experiment, although there are some exceptions such as La$_2$Fe$_{14}$B. For 1\,:\,5 compositions, a similar trend is evident.

\section{Machine Learning}
\subsection{Classical machine learning algorithms}

Machine learning (\gls{ml}) encompasses a variety of advanced statistical techniques. It creates a correspondence between a space of independent variables, $X$, and a space of dependent variables, $Y$, by taking a ground truth function $f_o : X_o \rightarrow Y$ representing a sequence of observations on a limited subset $X_o \subset X$ of the data to construct a more general function $f:X\rightarrow Y$ which extends the function $f_o$.  This ground truth function may also be represented as a list of observations $\{x_i\mapsto y_i\}_i\subset X_o \times Y$. The method by which $f$ is constructed from $f_o$ is referred to as \gls{ml}, and it is written symbolically as $f = ML(f_o) = ML(\{x_i\mapsto y_i\}_i)$, where \gls{ml} is optionally subscripted to indicate the algorithm or hyperparameters in use.

The goal of \gls{ml} is to represent the structure underlying $f_o$ as $f$. This is usually quantified by splitting the observations into a training set $f_{o}'$ and testing set $f_{o}^*$ so that $f_o' \cap f_o^* = \emptyset$. Then the function is constructed relative to a norm, termed a loss function, so that $||f_o' - ML(f_o')||$ is minimized in some appropriately defined subspace of $X\rightarrow Y$ to avoid overfitting. The quality of the fit with $f= ML(f_o')$ is measured by evaluating $||f_o^* - f||$ using the same class of norms. In this paper, all \gls{ml} algorithms use the standard Euclidean norm.

We use the \texttt{scikit-learn} library for the Python programming language for predictive data analysis \cite{pedregosa2011scikit}. The dataset (both experimental and theoretical) is split into a 70\,:\,30 train-to-test data set ratio for the model training. Performance does not change drastically by modifying the split ratio to 80\,:\,20. 
The classical linear \gls{ml} models used are:
\begin{enumerate}
\item Linear regresssion;
\item Lasso regularization;
\item Ridge regularization.
\end{enumerate}
The non-linear classical \gls{ml} models used are:
\begin{enumerate}
\item  Decision tree (DT) regression;
\item Random forest (RF) regression;
\item Gradient boost (GB) regression;
\item XGBoost (XGB) regression.
\end{enumerate}
In \gls{ml} training with cross-validation, the training dataset is further divided into two parts - the training and the validation datasets. Most \gls{ml} models have arbitrary hyperparameters, which are not modified directly during training. The validation dataset allows a meta-training of the \gls{ml} models by random or grid search on the training data and evaluation on the validation set without contaminating the testing set by predicting an \gls{ml} hyperparameter choice for test set evaluation. The most relevant cross-validation method is k-fold validation. The dataset is randomly split into k disjoint subsets, then one set is chosen as the validation dataset, while the others are chosen as training datasets.

\subsection{Artificial neural networks}

An artificial neural network
(\gls{ann}) is suitable for high complexity problems such as \gls{hc} prediction.
\begin{figure}
\centering
\includegraphics[scale=0.375]{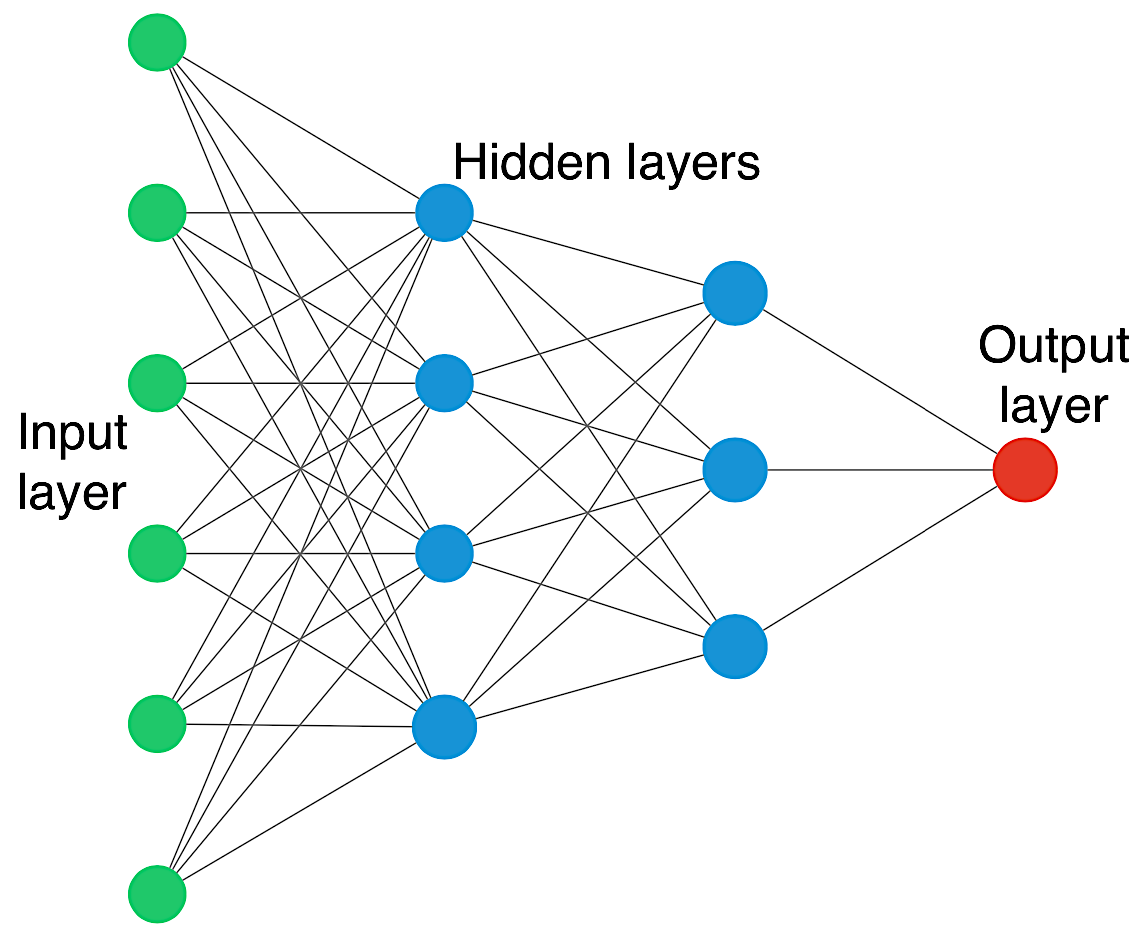}   \caption{Schematic diagram showing an artificial  neural network (only a portion used in the calculations) with 6 neurons in the input layer, 4 and 3 neurons in the second and the third deep (hidden) layers, and a neuron for regression in the output layer. Two different \gls{ann}'s were used in actual calculations, as discussed in the main text.}
    \label{fig:ANN}
\end{figure}
Fig.~\ref{fig:ANN} shows a schematic diagram (a portion used in the calculations) for an \gls{ann}  consisting of an input layer (6 neurons), two hidden layers (containing 4 and 3 neurons respectively), and the output layer with one neuron. A non-linear activation function \texttt{leaky-relu} [$f(z)=\max(0,z) + 0.01*\min(0, z)$] is used between the input and hidden layers to counter typical convergence problems found in small ANNs. We used the state of the art gradient descent benchmark optimizer Adam\cite{kingma2017adam}, as implemented in \texttt{tensorflow}\cite{tensorflow2015-whitepaper} in the Keras API\cite{chollet2015keras}.

\subsection{Performance comparison}

For the regression model, given an \gls{ml} fit, we use two common statistical measures of error applied to the $H_c$ predictions: (i) Mean Square Error (MSE):
\begin{equation}
{\rm MSE}=\frac{1}{N}\sum_i^N (y_i-y_{{pred}_i})^2,
\end{equation} which is also used for training all \gls{ml} models, and
(ii) R-squared ($R^2$):
\begin{equation}
    R^2 = 1-\frac{MSE}{V(y)},
    \label{Eq.R2}
\end{equation}
where $V(y)$ is the variance written as
\begin{equation}
V(y)=\langle(y_i-\langle y \rangle)^2\rangle
\end{equation}
with $\langle y\rangle$ as the mean of the predicted values. Equivalently, we may write $R^2=1-RSS/TSS$, where RSS is the Residual Sum of Squares $RSS=\sum_i^N (y_i-y_{{pred}_i})^2$, and $TSS$ is the Total Sum of Squares $TSS=\sum_i^N(y_i-\langle y \rangle)^2$. The loss function is usually written as a sum of the MSE and the first or second-order norms of the target function parameters. This provides convergence of the model, and penalizes overfitting for parameterized machine learning.

\subsection{Results and discussion}

\subsubsection{Machine learning on experimental data}

\begin{figure}
\includegraphics[scale=0.55]{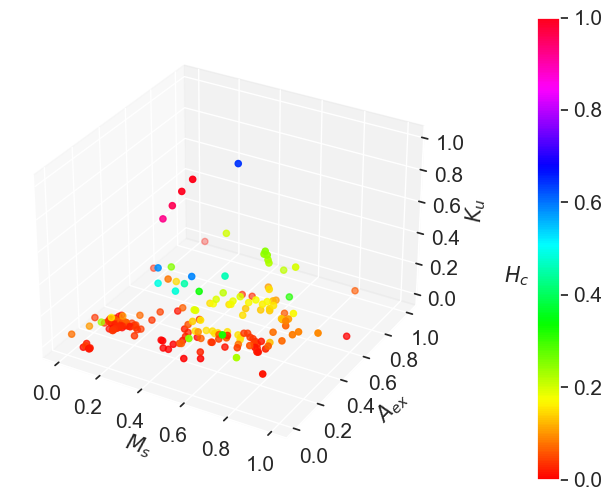}
\caption{The distribution of \gls{hc} as a function of \gls{ms}, \gls{aex}, and \gls{ku} in experimental magnetic materials. The physical quantities are rendered as normalized and dimensionless. The color map shows the magnitude of \gls{hc}.} 
\label{fig:exp_mats}
\end{figure}

\begin{table}
\begin{ruledtabular}
\caption{Comparison of $R^2$ and \gls{mse} metrics in different \gls{ml} models on test data sets for experimental materials. The non-linear regression models show better values of $R^2$ and \gls{mse} than the linear models. The error is evaluated over the difference of the logarithm of the coercivity in T.}
\begin{tabular}{l | l  l  }
Model & $R^2$ & \gls{mse}\\
\hline
Linear regression & 0.62& 0.64 \\
Lasso &0.38 & 1.07\\
Ridge &0.63 & 0.64 \\
Lasso-CV& 0.61&0.68 \\
Ridge-CV & 0.63&0.66 \\
Elasticnet & 0.62& 0.66\\
\gls{dt} regressor &0.89 & 0.19 \\
\gls{dt} pruned & 0.84 & 0.27\\
\gls{rf} regressor &0.87 &0.23\\
\gls{gb} regressor & 0.80&0.34\\
Tuned \gls{gb} regressor & 0.87 & 0.22\\
\gls{xgb} regressor & 0.87 & 0.23\\
Tuned \gls{xgb} regressor & 0.89 & 0.18\\
 \gls{ann} (Adam) & 0.64 &0.62\\
 \gls{lgbm}& 0.70& 0.60\\
 Fine tuned  \gls{ann} & 0.85 & 0.25
\end{tabular}
\label{metrics}
\end{ruledtabular}
\end{table}
Here, we explore the experimental data and the relation between target and independent variables. Fig.~\ref{fig:exp_mats} visualizes \gls{hc} as a function of \gls{ku}, \gls{ms}, and \gls{aex}. The actual values of the variables differ by several orders of magnitude. They are scaled using the \texttt{sci-kit} \texttt{minmaxscalar} in which a dimensionless scaled feature variable $x_{scaled}$ in the range [0,1] is obtained using the following relation:
\begin{equation}
x_{scaled} = \dfrac{(x - x_{min})} {(x_{max} - x_{min})},
\end{equation}
where $x_{min}$ and $x_{max}$ are minimum and maximum values of the feature variable $x$.
Fig.~\ref{fig:exp_mats} shows an increase of \gls{hc} with \gls{ku}, a decrease with \gls{ms}, and very little variation with \gls{aex}. 

In gradient-based algorithms, such as \gls{ann}, additive relationships are easier to model than multiplicative ones, so the independent and dependent variables are logarithmically scaled. Accordingly, the measures of model performance are given in terms of the logarithm of the \gls{hc}. As a result of this change, the absolute measure of $R^2$ is increased by $\sim 0.01$ for all gradient-based models.

\begin{figure} 
\centering
\includegraphics[scale=0.45]{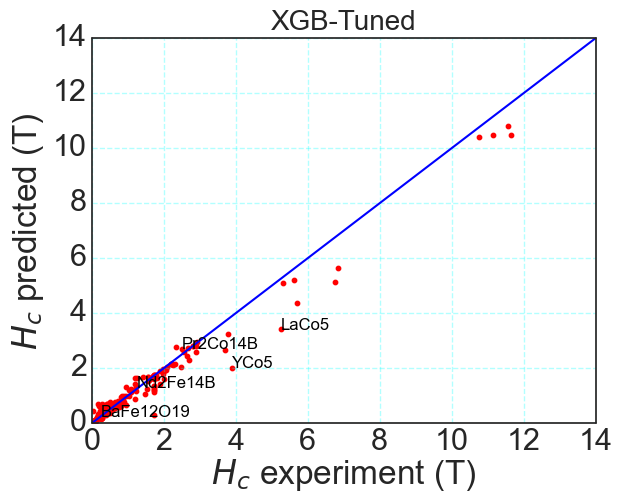}
\includegraphics[scale=0.45]{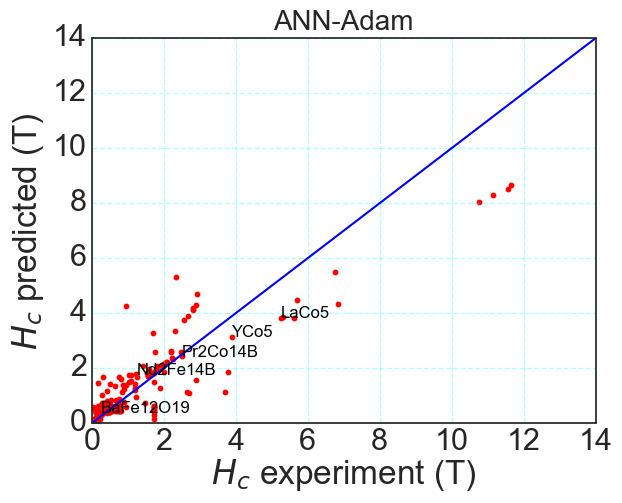}
\caption{ML predicted and experimental coercivities of magnetic materials obtained with tuned \gls{xgb} ({\it top}) and \gls{ann} ({\it bottom}). Selected key materials are labeled in the figure. Non-linear regressors all yield similar results.}\label{fig:ML}
\end{figure}

Figure \ref{fig:ML} shows a comparison of ML and experimental $H_c$ for  the test dataset. The XGB-tuned model shows better performance than ANN-Adam.
Table \ref{metrics} shows the performance of various \gls{ml} models. The linear models show poor performance, with $R^2\ll 1$. This demonstrates the complex non-linear relation between the dependent and independent variables. The non-linear models \gls{dt}, \gls{xgb}, and \gls{rf} perform better, with $R^2$ values of $0.89$, $0.87$, and $0.87$, respectively. These values still witness a decrease from the $R^2$ scores of more than $0.9$ on the training dataset.

The \gls{ann} model has $R^2$ scores that are similar to other classical models (see Table \ref{metrics}). The \gls{ann} architecture using (dense layers=64, activation function=leaky\_relu, dense layers=32, activation function=leaky\_relu, dense layers=16, activation function=leaky\_relu, and output layer=1, and learning rate=0.007, epochs=1000, batch size=24) showed similar $R^2$ scores to \gls{rf}. This weak performance for the \gls{ann} architecture is attributed to the smallness of  the dataset size.

\paragraph*{Feature importance.}

The model performance can be further validated from the dependence of \gls{hc} on the independent features.
We compute the importance of \gls{rf} features by utilizing the model-agnostic interpretive features of \texttt{scikit-learn} on the \gls{rf} and \gls{xgb} regression models \cite{BreiFrieStonOlsh84}. The feature importance is computed 
by the Gini index 
\begin{equation}
    {\rm Gini~index} = 1-\sum_i p_i^2\,,
\end{equation}
where $p_i$ is the probability of class $i$ in the data. 
\begin{figure}
\centering
\includegraphics[scale=0.45]{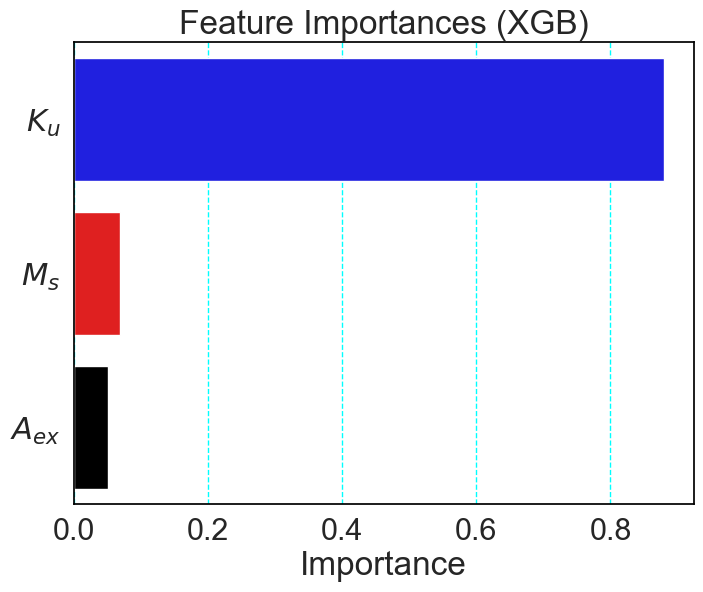}
\caption{Histogram of important features contributing to \gls{hc} in the \gls{xgb} model. The \gls{ku} clearly leads, followed distantly by \gls{ms} and \gls{aex}.}
    \label{fig:importance}
\end{figure}
Fig.~\ref{fig:importance} depicts the relative importance of the independent parameters for \gls{hc}, as computed by the \gls{xgb} model. \gls{ml} accurately identifies \gls{ku} as the leading contributing feature to \gls{hc}. More surprisingly, the effect of \gls{ms} is about four times smaller than that of \gls{ku}.  
The uneven distribution of magnetic materials (shown in Fig.~\ref{fig:exp_mats}), and a constant noise level, may explain the unequal weighting of $K_u$ and $M_s$. Finally, we note that \gls{aex} has a comparatively minimal effect on \gls{hc}.


\subsubsection{Machine learning prediction from micromagnetic data}

\begin{figure}
\centering
\includegraphics[scale=0.525]{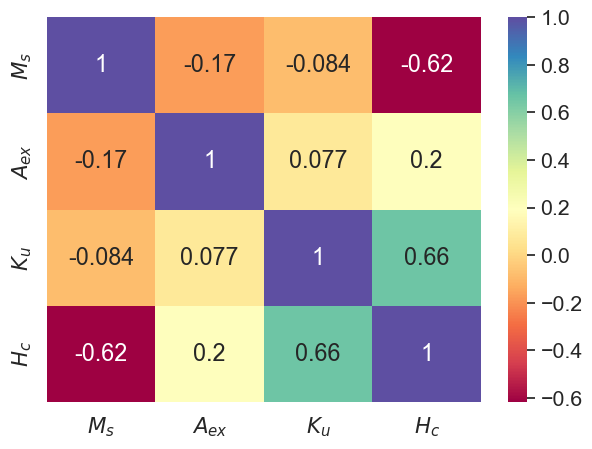}
\caption{Heat map showing the correlation between the \gls{hc}, \gls{aex}, \gls{ku}, and \gls{hc}. The color palette corresponds to the correlations between the variables. \gls{hc} is strongly correlated with \gls{ku}, negatively correlated with \gls{ms}, and weakly correlated with \gls{aex}. \gls{ms} shows a slightly negative correlation with \gls{ku} and \gls{aex}. \gls{ku} shows a slightly positive correlation with \gls{aex}.}
    \label{fig:heatmap}
\end{figure}

\begin{figure}
\centering
\includegraphics[scale=0.575]{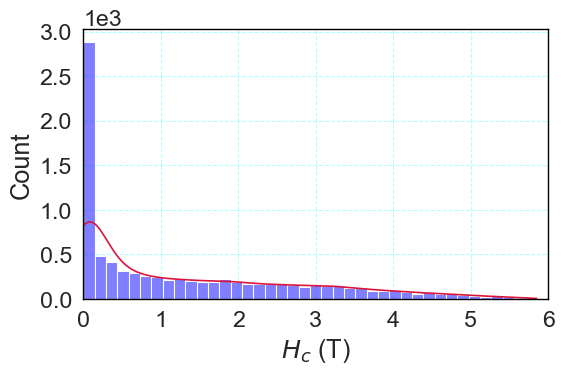}
\caption{Distribution of \gls{hc} in the training set of 8770 data points. The $y$-axis is scaled by $10^3$. The data points are right-skewed.}
\label{distplot}
\end{figure}

In order to explore the relation between \gls{hc} and the independent variables, we used \TheorySize ~micromagnetic simulation-generated data points as a training set. The correlation between variables is given by Pearson’s correlation coefficient, which is defined as
\begin{eqnarray}
r=\dfrac{\sum(X_1-\langle X_1\rangle)(X_2-\langle X_2\rangle)}{\sqrt{\sum(X_1-\langle X_1\rangle)^2\sum (X_2-\langle X_2\rangle)^2}},
\end{eqnarray} 
where $X_i$ and $\langle X_i\rangle$ denote the variable $i$ and its average value. A positive (or negative) value of $r$ indicates a positive (or negative) correlation between the two variables: the higher the $r$-value, the higher the correlation. Fig.~\ref{fig:heatmap} shows a correlation heatmap for the variables. The distribution for the \TheorySize-data set is right-skewed, as shown in Fig.~\ref{distplot}, with a peak around 0.2--0.4 T. It is similar to the distribution for the experimental materials (not shown here): most of the materials have values around the peak, and a few materials, such as 1\,:\,5 compositions, have larger values of 2--6  T. For better \gls{ml} training, logarithmic transformation is appropriate for skewed data, which is employed in our model fits.

Significantly, the strongest (absolute) correlation is observed between \gls{hc} and \gls{ku} at 0.66, followed by \gls{ms} with a correlation of $-0.62$, and then by \gls{aex} at 0.2. This pattern suggests a high dependence of \gls{hc} on \gls{ku}. The negative correlation between \gls{hc} and \gls{ms} aligns with the theoretical relationship $H_c \propto M_s^{-1}$. Similarly, the correlation coefficient ($r$) between \gls{hc} and $K_u$ or $A_{ex}$ validates the proportionalities $H_c \propto K_{u}$ and $H_c \propto A_{ex}$. These findings are consistent with the trends observed in actual materials, as illustrated in Fig.~\ref{fig:exp_mats}. 

Next, we discuss the model performance in micromagnetic simulation data, as given in Table~\ref{metricsmumax}. Both linear and non-linear models perform well, as the $R^2$ score is above $0.9$. This is expected in a linear regression model where multivariate analysis demonstrates very little multicollinearity in randomly generated independent variables. The multicollinearity can be measured with the
\gls{vif} $VIF_{\alpha}=(1-R_{\alpha}^2)^{-1}$, where $R_{\alpha}^2$ is the $R^2$ [Eq. (\ref{Eq.R2})] value of \gls{hc} considered as a function of variable $\alpha$. The computed \gls{vif} factors are $1.53$ for \gls{ms}, $4.36$ for \gls{aex}, and $1.53$ for \gls{ku}.

\begin{figure}
    \centering
    \includegraphics[scale=0.7]{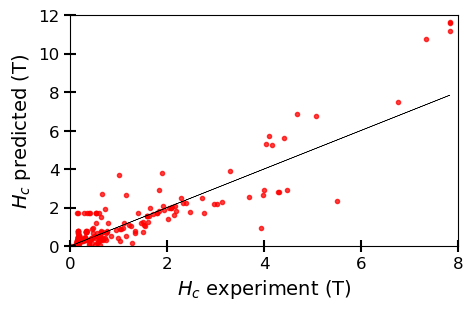}
    \caption{Scatter plot showing the comparison of experimental and ML-mumax$^3$ (ML-micromagntics data) predicted $H_c$ for  experimental materials.}
    \label{expvsML}
\end{figure}

More advanced decision tree regressors improve the performance. \gls{xgb}, as the most accurate model, was used as a benchmark, with the hyperparameters given in Appendix~A for \gls{hc} prediction from experimental data. The deviance in the \gls{mse} of the training and testing data sets is shown in Fig.~\ref{fig:deviance}. \gls{ann}s exhibit performance similar to other non-linear regressors. The ANN is as follows: (dense layers=64, activation function=leaky\_relu, dense layers=32, activation function=leaky\_relu, dense layers=16, activation function=leaky\_relu, and output layer=1, and learning rate=0.007, epochs=1000, batch size=56).
To obtain a deeper insight into \gls{ml} predictability, the experimental,  scaled-micromagnetic, and  \gls{ml}-predicted \gls{hc} values are compared with only the known important rare-earth-based materials in Table~\ref{mumaxMLprediction}. In general, \gls{ml}-predicted values are smaller than the micromagnetically computed  values. For 2\,:\,14\,:\,1 rare-earth magnets, \gls{ml} improves the results. For 1\,:\,5 rare-earth magnets, it slightly underestimates the experiment. The full comparison of experimental and ML predicted data is given in Fig. \ref{expvsML}

Additionally, \gls{xgb} was trained with different fractions of the dataset, from 200 to 12000, by computing the \gls{mae} error metric: 
\begin{equation}
    {\rm MAE} = \sum_i^N \dfrac{|y_i-y_{pred_i}|}{N}\,
\end{equation}

\begin{figure}
\centering
\includegraphics[scale=0.575]{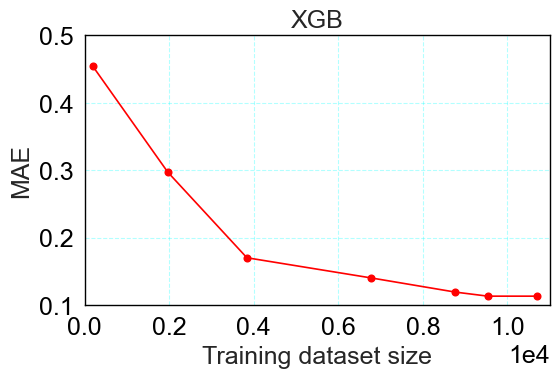}
\caption{\gls{mae} convergence for \gls{hc} with respect to training data set size in \gls{xgb} model. The $x$-axis is in the scale of $10^4$.}
\label{fig:trainsize}
\end{figure}
Fig.~\ref{fig:trainsize} shows the convergence of the \gls{mae} (obtained with the \gls{xgb} model) plotted against the training data set size, indicating that the model performs well for datasets with more than 6000 points.

\begin{table}
\begin{ruledtabular}
\caption{\mumax data on hypothetical materials: Comparison of $R^2$ and \gls{mse} of \gls{hc} for different \gls{ml} models. All the errors are evaluated over the $\ln$ of the \gls{hc} in T. The fine-tuned model hyperparameters are given in Appendix A.}
\begin{tabular}{l |  c  c  }
Model & $R^2$ & $MSE$\\
\hline
Linear regression	& 0.95	&0.22\\
Lasso	&0.54	&2.13\\
Ridge	&0.95	&0.22\\
Lasso-cv	&0.95	&0.25\\
Ridge-cv	&0.95	&0.22\\
Elasticnet	&0.94	&0.26\\
\gls{dt} regressor	&0.97	&0.13\\
\gls{dt} pruned	&0.97	&0.15\\
\gls{rf} regressor	&0.98	&0.08\\
\gls{gb} regressor	&0.98	&0.09\\
Tuned \gls{gb} regressor	&0.98	&0.08\\
\gls{xgb} regressor	&0.98	&0.98\\
Tuned \gls{xgb} regressor	&0.98	&0.08\\
\gls{ann}  & 0.98 & 0.09
\end{tabular}
\label{metricsmumax}
\end{ruledtabular}
\end{table}

Next, we discuss the loss function RMSE to examine the model performance.
We computed the \gls{rmse} as
\begin{equation}
{\rm RMSE} =\sqrt{\frac{1}{N} \sum_{i=1}^N(y_i - \hat{y}_i)^2}\,,
\end{equation}
where $y$ and $\hat{y}$ refer to the experimental and ML/\mumax predicted values.
Remarkably, the \gls{rmse} and MAE between experimental and \gls{ml} (tuned XGB) predicted values are smaller than the \gls{rmse} between experimental and the scaled-\mumax, indicating that \gls{ml} helps to correct the \mumax predictions. It also shows how the \gls{hc} is correlated with other features. Moreover, \gls{ml} is an order of magnitude computationally faster than \mumax, and we can use the trained model for predicting \gls{hc} in new materials without any \mumax calculations.

\begin{table}
\begin{ruledtabular}
\caption{\gls{hc} (in T) comparisons among experiment, \gls{ml} prediction, and scaled \mumax for different permanent magnetic materials. The \gls{ml} (tuned \gls{xgb} and  \gls{ann}) predictions are obtained by training the hypothetical materials using the random independent variables \gls{ms}, \gls{aex}, and \gls{ku}, and the \mumax computed \gls{hc}. \gls{rmse} and \gls{mae} quantify the discrepancy between the experimental \gls{ml} and the \mumax scaled $cH_c$.}
\begin{tabular}{l | ll  c l }
Material & \gls{hc}(exp) & \gls{hc}(\gls{xgb}) & \gls{hc}(\gls{ann} ) &$cH_c$\\
\hline
La$_2$Fe$_{14}$B	& 0.52	&0.52&0.48 &0.53 \\
Ce$_2$Fe$_{14}$B	&0.54	&0.84	&0.73&0.79\\
Pr$_2$Fe$_{14}$B	&1.54	&1.59	&1.54&1.62\\
Nd$_2$Fe$_{14}$B	&1.24	&1.50	&1.45&1.56\\
Gd$_2$Fe$_{14}$B	&0.23	&0.56 &0.52&0.60\\

Tb$_2$Fe$_{14}$B	&2.93	&4.00	&5.65&5.41\\

Dy$_2$Fe$_{14}$B	& 1.70	&2.76	&3.80&3.70\\

Ho$_2$Fe$_{14}$B	&0.75	&1.50	&1.76&1.82\\

Lu$_2$Fe$_{14}$B	&0.29	&0.59	&0.52&0.58\\

Y$_2$Fe$_{14}$B	    &0.19	&0.57&0.48&0.54\\

Th$_2$Fe$_{14}$B	&0.17	&0.56	&0.48&0.53\\

La$_2$Co$_{14}$B	&0.34	&0.65 &0.68&0.77\\

Pr$_2$Co$_{14}$B	& 2.50	&2.28	&2.24&2.36\\

Nd$_2$Co$_{14}$B	&3.69 &1.00 &0.90 &1.06\\

Gd$_2$Co$_{14}$B	&0.30 &0.67	&2.79 &2.37\\

Y$_2$Co$_{14}$B	    &0.34	&0.62	&0.61&0.71\\

NdLaCeF$_{14}$B      & 0.65	&1.33  &1.25& 1.39\\
LaCeYFe$_{14}$B	&0.41	&0.99	&1.26&1.39	\\																	
NdPrFe$_{14}$B	&0.80	&1.33	& 1.27&1.39\\																	
Sm$_2$Co$_{17}$	&1.25	&1.60	& 1.53&1.62		\\																
Sm$_2$Fe$_{17}$N$_3$	&2.30	&2.46	& 3.13 &2.95\\																	
SmCo$_5$ &	7.50 &6.76 & 6.66 &9.70\\
YCo$_5$	& 3.90	&2.63	&2.68  &3.33\\
LaCo$_5$	& 5.25	&4.18	& 4.14 &4.22\\
CeCo$_5$	& 5.70	&4.10	&5.10  &5.17\\
PrCo$_5$	& 5.32	&4.05	&3.81  &4.02\\
NdCo$_5$	&0.15	&0.12	&0.10 & 0.12 \\
\hline
\gls{rmse} & & 0.90& 1.10& 1.10\\
\gls{mae} & &0.68 & 0.78 & 0.80\\
\end{tabular}
\label{mumaxMLprediction}
\end{ruledtabular}
\end{table}

\section{ML application with DFT}

Our computational  \gls{hc} predictions involve two steps: \gls{dft} for Ce-doped Nd$_2$Fe$_{14}$B, and then \gls{ml} trained on micromagnetically generated databases with input parameters from the DFT computations.

\subsection{DFT methods and crystal structure}

\begin{figure}
\centering
\includegraphics[scale=0.275]{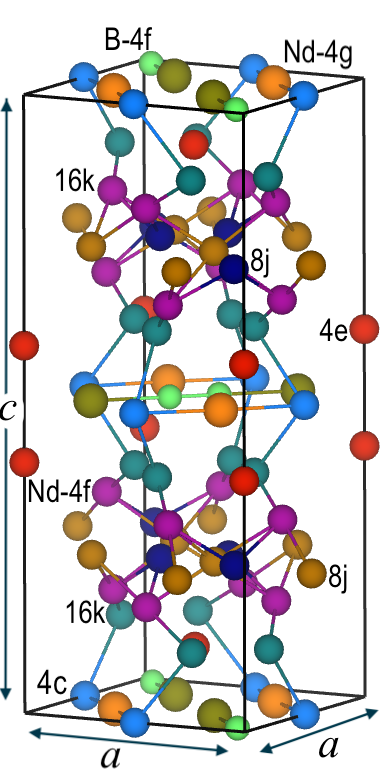}
\caption{Crystal structure of Nd$_2$Fe$_{14}$B occupied by Nd at two non-equivalent sites $4f~{\rm and} ~4g$, Fe at six inequivalent sites $16k,~ 16k,~8j,~8j,~4c,~{\rm and}~4e$ (labeled by Wyckoff positions), and B at the $4f$ site.}\label{crystalstructure}
\end{figure}

We used the \gls{vasp} \cite{vasp1,vasp2} with the \gls{paw} formalism in the \gls{gga} of \gls{pbe} semi-local exchange-correlation functionals\cite{PBE,PBE1} including onsite electron-electron correlation Dudarev-Hubbard\cite{Dudarev} $U_{\mathrm{eff}} =U -J$ for the $4f$ states of Nd, and spin-orbit interaction. The kinetic energy cut-off is 520 eV for the plane wave expansion, with a $4\times4\times2$ {\bf k}-mesh for the Brillouin zone integration. The experimental structure was used in calculations to avoid overestimating bond lengths lengths and lattice constants in the \gls{pbe}.

The crystal structure of Nd$_2$Fe$_{14}$B is tetragonal with 68 atoms (4 formula units), having space group $P42/mnm$ (136) and lattice constants $a=b=8.80$ and $c=12.20$ \AA, taking $\alpha=\beta=\gamma=90^\circ$ at a temperature $\mathrm{T}=285\, ~\mathrm{K}$\cite{HerbstRevModPhys91,IsnardJAP95}, with experimental parameters for full Ce-doped Nd$_2$Fe$_{14}$B\cite{OesterreicherMRB84}.
The crystal structure of the 2\,:\,14\,:\,1 neo-magnet is shown in Fig.~\ref{crystalstructure}, which consists of rare-earth Nd in the 2 inequivalent sites $4f$ and $4g$, transition element Fe in the 6 inequivalent sites  at $16k$, $16k$, $8j$, $8j$, $4e$, and $4c$, and finally B at the $4f$ sites. Fully self-consistent spin-orbit calculations, i.e.,  \gls{pbe}+$U$+\gls{soc},  were performed to obtain the magnetic anisotropy energy along with the spin and orbital magnetic moments.

 \subsection{Magnetic properties}
Here we discuss the \gls{pbe}+$U$+\gls{soc} results with Nd$_2$Fe$_{14}$B, using \gls{ml} for $H_c$ prediction. Hubbard $U$-correction is necessary for highly correlated $3d$ and $4f$ elements\cite{DFTU,BhandariPRA016, BhandariPRM021, BhandariPRR022} to accurately predict $K_u$. We performed test calculations for various values of $U$ from 4--7 eV for Nd $4f$ orbitals. Table~\ref{moments} shows the spin and orbital magnetic moments of  Nd, Fe, and B at different crystallographic sites for different values of $U_{\mathrm{eff}}$. The non-magnetic B atom carries negligible magnetic moments. Nd exhibits an Nd$^{3+}$ state with spin magnetic moment ($\mu_s$) of $\sim -3.3$ and strongly quenched orbital moment ($\mu_l$) of $\sim 1.5$ $\mu_B$ due to the crystalline electric field. The spin and orbital moments have opposite signs, consistent with Hund's rule for a less than half-filled $4f$ shell. The net magnetic moment is robust, and does not vary significantly with $U$.

\renewcommand\arraystretch{1.5}
\begin{table*}
\begin{ruledtabular}
\caption{Spin and orbital magnetic moments ($\mu_s$ and $\mu_l$) of individual atoms in different Wyckoff positions, and total magnetic spin and orbital magnetic moments ($\mu_{st}$ and $\mu_{lt}$) per unit cell (in $\mu_B$), in ${\rm Nd_2Fe_{14}B}$ computed using the \gls{pbe}+$U$+\gls{soc} method with various values of $U$ for the Nd ($4f$) electronic states. There are two $16k$ and $8j$ sites, therefore, the pairs are given. The $\mu_l$ of the pair at $8j$-sites do not differ, and only one value is given to represent both atoms.}\label{moments}
\begin{tabular}{c | ll |ll| ll | ll | ll | ll | ll |ll}
atom$\rightarrow$ &\multicolumn{2}{c|}{Nd($4f$)} &
\multicolumn{2}{c|}{Nd($4g$)} & \multicolumn{2}{c|}{Fe($16k$)} &  \multicolumn{2}{c|}{Fe($8j$)} & \multicolumn{2}{c|}{Fe($4e$)} & \multicolumn{2}{c|}{Fe($4c$)} & \multicolumn{2}{c|}{B($4g$)} & total
\\\cline{2-17}
$U\downarrow$&$\mu_s$ & $\mu_l$ &$\mu_s$ &$\mu_l$  &$\mu_s$& $\mu_l$    &$\mu_s$&$\mu_l$ &$\mu_s$& $\mu_l$& $\mu_s$&$\mu_l$  & $\mu_s$& $\mu_l$ & $\mu_{st}$& $\mu_{lt}$
\\
    \hline
4 eV   & -3.26 &  1.52 &-3.28  &1.53 &2.27,2.36  &0.043,0.049 &2.29,2.71 & 0.044 & 2.03& 0.045& 2.48 & 0.054 &-0.17 &0.00 & 105.24 & 14.75\\

5 eV   & -3.26 & 1.52 & -3.28 & 1.53& 2.27,2.36 & 0.042,0.048&  2.29,2.70& 0.044&2.03&0.045 &2.47&0.052&-0.17 &0.00 &105.25 & 14.75\\

7 eV  &-3.27  &1.52  &-3.28 &1.53 &2.27,2.36  &0.042,0.048 & 2.29,2.71 & 0.042& 2.03&0.044 &2.48 &0.049 &-0.17 &0.00 &105.25 & 14.68\\
\end{tabular}
\end{ruledtabular}
\end{table*}

Table~\ref{magneticaniso} lists  $K_u$, $M_s$,  $T_{\rm C}$, and their comparison with experimental values. $K_u$ is positively correlated with $U$. The experimental $K_u$ is 4.5 MJ/m$^3$\cite{Sagawa} at 300 K, which corresponds to the computed value for $U\sim 4$ eV. 
$T_{\rm C}$ is computed with the static Green's function (GF) as implemented in \gls{asa} in the \gls{lda}\cite{LDA} for the exchange-correlation functional in the \gls{lmto} program\cite{PASHOV2020107065, Schilfgaarde99}. The pair exchange interaction $J_{RR^{'}}$ between magnetic $R$ and $R^{'}$ ions is computed using the Lichtenstein formula\cite{Liechtenstein87}, from which $T_{\rm C}$ is estimated.
The calculated Weiss mean-field theory\cite{WeissMF} value of 718.1 K is larger than the experimental value, which is expected in the \gls{mfa}. According to a spin-wave theory by Tyablikov\cite{tiablikov2013methods}, the \gls{rpa} corrects the value, bringing it down to 539.1 K. The experimental value lies within the \gls{mfa} and \gls{rpa} limits.

\begin{table} 
\begin{ruledtabular}
\caption{Calculated \gls{ku} in MJ/m$^3$, \gls{ms} in $\mu_B$/f.u. using PBE$+U$+SOC, \gls{ct} in Kelvin (K) using Green's function-ASA (\gls{lda}), and comparison with available experiment.}\label{magneticaniso}
\begin{tabular}{c|c | c |c |c | c}
$U$ & $K_u$ &  $M_s$ & $M_s$(Expt \cite{HerbstRevModPhys91}) & $T_{\rm C}$ & $T_{\rm C}$(Expt\cite{HerbstRevModPhys91}) \\
\hline
4 eV     & 6.16 &29.98 &37.7\footnote{at low temperature, (4 K)}, 32.5\footnote{at high temperature (295 K)}  &718.1\footnote{Mean Field Approximation (MFA)}, 539.1\footnote{Random Phase Approximation (RPA)} &585\\
5 eV     & 8.56 &30.00 &$\cdots$  &  & $\cdots$\\
7 eV     &11.58  &30.01 & $\cdots$ &  &$\cdots$\\
\end{tabular}
\end{ruledtabular}
\end{table}

Next, we discuss the site-resolved spin-orbit anisotropy energy which is computed as $E_{\mathrm{anis}}=E_{100}-E_{001}$\cite{BhandariPRA016}, where $E_{100} ~\rm{and}~ E_{001}$ are PBE+$U$+SOC-computed atomic site energies for spin-quantization along the [100] and [001] directions.
Generally, $4f$-elements contribute to \gls{ku}, and $3d$-elements contribute to the magnetic moments in rare-earth-based magnets. The spin moments of the rare-earth ion are anti-parallel to the spin moment of the $3d$ ion, which is also the case for neo-magnets. The site contribution of the Nd-element to crystalline anisotropy energy is given in Table~\ref{tab:sitecontribution}. In our \gls{pbe}+$U$+\gls{soc} calculations, we did not impose symmetry, which means that all the atoms are inequivalent under the $P_1$ crystal symmetry. All eight Nd atoms split into eight different sites.
The individual on-site energies differ significantly for the atoms of the same crystallographic site. Interestingly,  at least one Nd (Nd$_1$) at $4f$ has a negative contribution to crystalline anisotropy energy, as given in Table~\ref{tab:sitecontribution}. Theoretically, it can be inferred that Nd has a tendency to be planar at the $4f$ site, differing from the $4g$ site, which is strictly uniaxial along the crystalline $ c$ direction. These results are consistent with experiment\cite{HaskelPRL05}.

\begin{table}
\caption{Site resolved spin-orbit anisotropy energy ($E_{\mathrm{anis}}$) in meV of the Nd contribution to magneto-crystalline anisotropy energy. In PBE+$U$+SOC calculations, the crystalline symmetry lowers to the $P_1$ point group implying that the atoms (including the Nd) split into eight sites.
The negative value for Nd at the $4f$ site indicates it produces a planar contribution to the crystalline anisotropy energy, consistent with  experiment\cite{HaskelPRL05}.}\label{tab:sitecontribution}
\begin{ruledtabular}
\begin{tabular}{l c|l}
Atom &  Wyckoff-position & $E_{aniso}$\\
\hline
Nd$_1$ & ${4f}$ &  ~1.3780 \\
Nd$_2$ & ${4f}$ & -1.1914\\
Nd$_3$ & ${4f}$ & ~0.5336\\
Nd$_4$ & ${4f}$ & ~6.2204\\
Nd$_5$ & ${4g}$ & ~0.4865\\
Nd$_6$ & ${4g}$ & ~3.8064\\
Nd$_7$ & ${4g}$ & ~6.2094\\
Nd$_8$ & ${4g}$ & ~3.0622\\
\end{tabular}
\end{ruledtabular}
\end{table}

\begin{table}
\caption{Total spin, orbital, and spin + orbital magnetic moments  $\mu_s$, $\mu_l$, and $\mu_s+\mu_l$ in $\mu_B$/cell, $K_u$ in MJ/m$^3$, and $M_s$ in MA/m of Ce$_2$Fe$_{14}$B and  Nd$_2$Fe$_{14}$B.}\label{cedoped}
\begin{ruledtabular}
\begin{tabular}{l|l|l}
Properties &Nd$_2$Fe$_{14}$B & Ce$_2$Fe$_{14}$B\\
\hline
  $\mu_s$ &105.24 & 120.73 \\
  $\mu_l$ &14.75  &4.55\\
  $\mu_s+\mu_l$ & 119.99 &125.28\\
  $K_u$ &6.16 &1.21 \\
  $M_{s}$ & 1.31 & 1.24
\end{tabular}
\end{ruledtabular}

\end{table}
Table~\ref{cedoped} shows the computed values of the magnetic moments and magnetic anisotropy in the Ce-substituted neo-magnet at Nd-$4f$ sites. The Ce-atom carries a small spin magnetic moment $\sim 1 \mu_B$ compared to the Nd $\sim 3\mu_B$ moment. Moreover, the orbital moment does not entirely cancel the spin moment in Nd, unlike in Ce, where the net spin + orbital moment vanishes. Therefore, there is an increase in the net magnetic moment with Ce.
On the other hand, magnetic anisotropy is reduced because the Ce-site contribution is much smaller than that of Nd. The calculated value of $K_u$ slightly underestimates the experiment (see Table~I), which is reasonable, given the choice of the $U$-values used for the Ce and Nd $4f$ states. We used $U_{\mathrm{eff}}=2$ eV for Ce and $4$~eV for Nd in the calculations.


\subsection{ML coercivity prediction}

\begin{figure}
\centering
\includegraphics[scale=0.35]{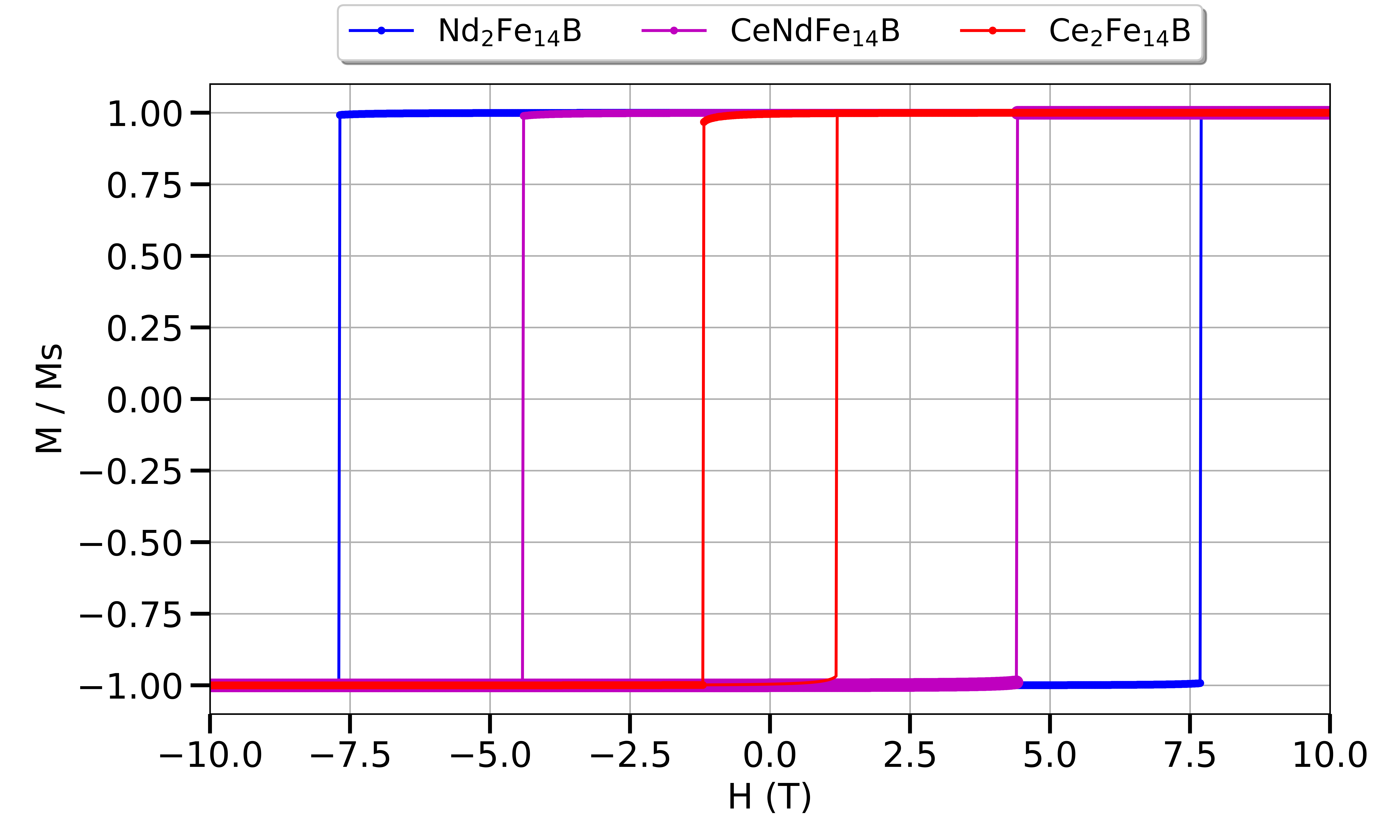}
\caption{Hysteresis curve for Nd$_{2-x}$Ce$x$Fe$_{14}$B ($x=0,~1.0, ~2.0$) calculated from micromagnetic simulation. Ce leads to the reduction in the coercivity. $M/M_s$ is the reduced magnetization.}
\label{fig:coercivity}
\end{figure}

\begin{figure}
\centering
\includegraphics[scale=0.6]{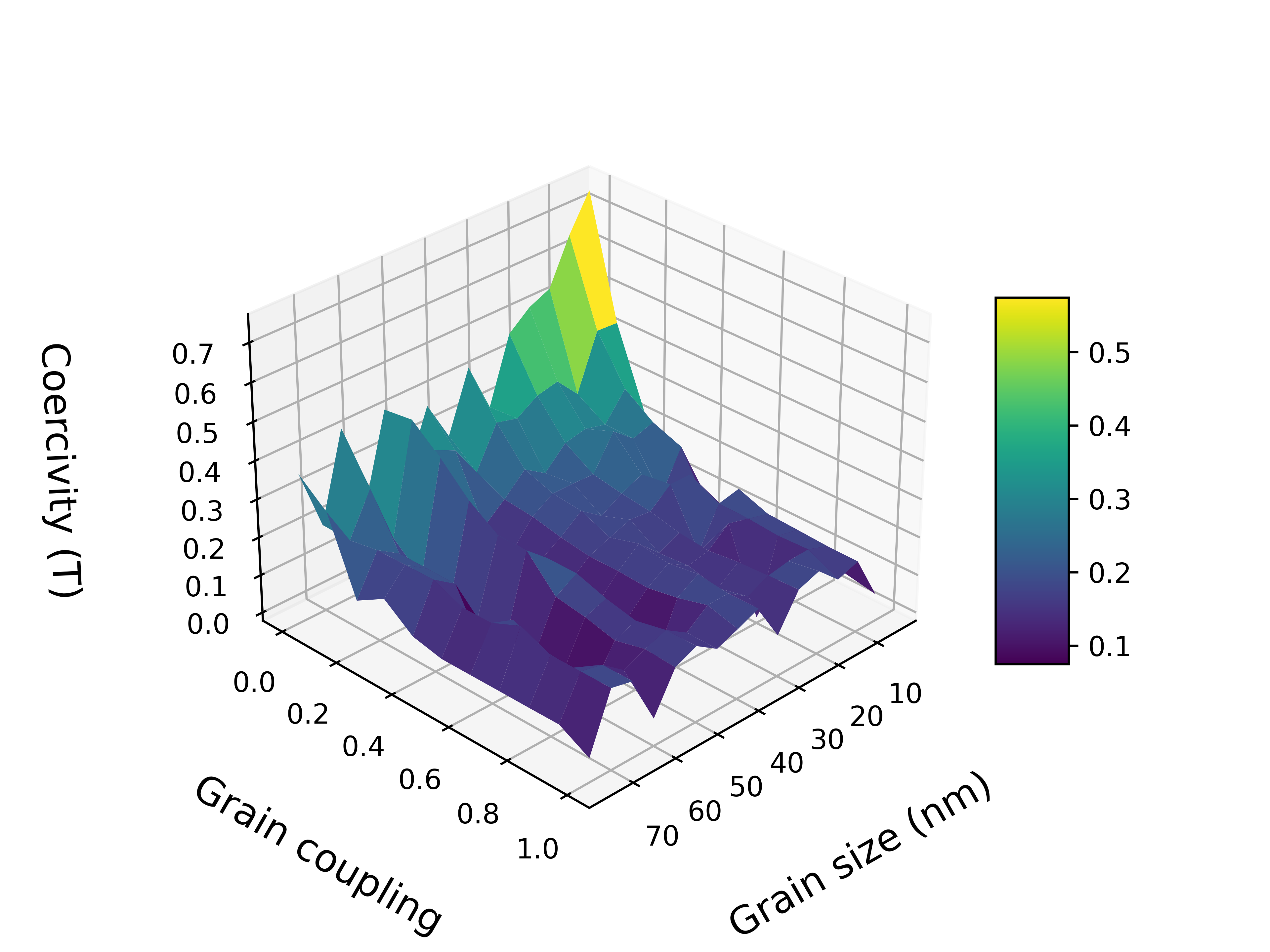}
\caption{Coercivities for different inter-grain couplings for Nd$_{2}$Fe$_{14}$B, measured as a function of the ratio of inter-grain exchange stiffness to intra-grain exchange stiffness and different grain sizes. As the material is complex, a cuboid of $128 \times 128 \times 128$ nm$^3$ was generated for each grain size, resulting in some random undulations according to grain initialization. The plot displays a decreasing coercivity with increasing grain size as observed in experiment \cite{PathakAdvacedMat015}.}
\label{fig:meta_coercivity}
\end{figure}

\begin{figure}
\centering
\includegraphics[scale=0.35]{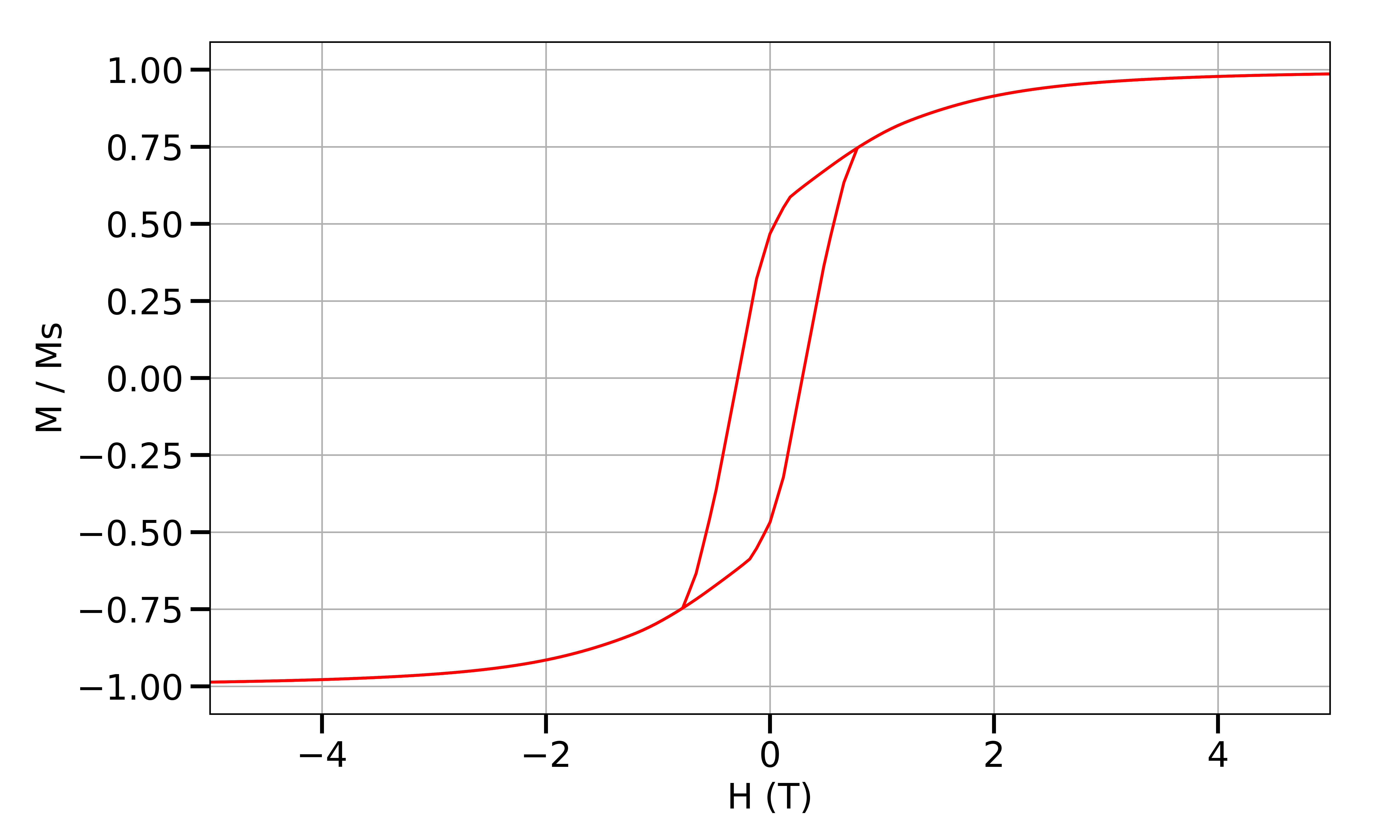}
\caption{Hysteresis curve for Ce$_2$Fe$_{14}$B  calculated naively with micromagnetic simulation with grain size set to $30$nm and intergrain stiffness to intragrain stiffness set at $0.58$ to recreate the \gls{ml}  \gls{hc} prediction.}
\label{fig:coercivity_mod}
\end{figure}

Central to this paper is the hypothesis that macroscopic measures of \gls{hc} are not well predicted by theory, while the microscopic magnetic parameters \gls{ms}, \gls{ku}, and \gls{aex} are. However, knowledge of the microscopic parameters is sufficient to produce realistic estimates of \gls{hc} by accumulating a database and modeling it with \gls{ml}. Furthermore, once an \gls{ml} prediction is made, information about the material grain structure may be reverse-engineered with micromagnetic modeling. Although material fabrication variations have a significant impact on \gls{hc}, this section demonstrates the surprising effectiveness of the methodology as  applied to Nd$_{2}$Fe$_{14}$B, with and without Ce doping. 



\begin{table}
\caption{Comparison of \gls{ml} predicted and scaled-micromagnetic ($cH_c$) \gls{hc} in T for materials with calculated DFT parameters. ML-exp/ML-mumax$^3$ represent to predictions made by training experimental/scaled-micromagnetic data.}\label{MLpred}
\begin{ruledtabular}
\begin{tabular}{l|l|l|l}
Material &  ML-Exp & ML-mumax$^3$ &$cH_c$\\
\hline
Nd$_2$Fe$_{14}$B & 3.92 & 2.39 &2.35\\
CeNdFe$_{14}$B & 0.88 & 1.27 &1.29\\
Ce$_2$Fe$_{14}$B & 0.39 & 0.40 & 0.36 \\
SmCo$_5$  & 8.87 & 5.88 & 18.12\\
YCo$_5$  & 1.45 & 1.51 & 1.60\\
CeZrFe$_{11}$ & 6.25 & 4.93 & 5.99\\
CeZrFe$_{11}$N &1.51 & 2.15 & 2.59 \\
CeTiFe$_{11}$ & 1.50 & 2.09 & 2.70 \\
CeTiFe$_{11}$N & 1.51 & 1.92 & 2.45\\
\end{tabular}
\end{ruledtabular}

\end{table}

The \textit{ab initio}-calculated values were fed to the tuned \gls{xgb} \gls{ml} model to produce the output predictions. Additionally, micromagnetic simulation was used to determine a hysteresis loop. The results are depicted in Table~\ref{MLpred}. As can be seen, the reduction trend of $H_c$ with Ce is similar to that found experimentally\cite{PathakAdvacedMat015}. Further, we computed the ML [trained with experiment (ML-Exp) and scaled micromagnetic data (ML-mumax$^3$) separately) and scaled micromagnetic simulated $H_c$ for SmCo$_5$, YCo$_5$, CeZrFe$_{11}$, CeZrFe$_{11}$N, CeTiFe$_{11}$, and CeTiFe$_{11}$N using DFT intrinsic parameters as given in Table~\ref{MLpred}. The ML-exp predicts inconsistent $H_c$ due to the small data size training, as evident from model performance metrics in Tab II. The \gls{ml}-mumax$^3$ trained on big data sets produces more sensible $H_c$ than scaled-micromagnetics and ML-exp.
For SmCo$_5$, micromagnetic simulation yielded a factor of three larger $H_c$ than the \gls{ml}-mumax$^3$ model and experiments. For YCo$_5$, the \gls{ml}-mumax3$^3$ model produced slightly lower $H_c$ than the experiment, attributed to the DFT underestimate of $K_1$. Overall the \gls{ml}-mumax$^3$ model consistently tends to correct the micromagnetically predicted $H_c$ towards experiments for known materials (see Table~\ref{tabcoercivity}).

If the fabricated material is grown as a crystal with well-defined grains, it is possible to reverse-engineer the structure. The dependence of \gls{hc} on micromagnetic parameters viz, grain sizes and inter-grain coupling is visualized in Fig.~\ref{fig:meta_coercivity} in $128\times 128\times 128$ nm$^3$ magnetic cell. Here the grain coupling refers to a reduction factor to the stiffness exchange coupling $A_{ex}$. We note that we did not find much difference in the results with the use of $64\times 64\times 64$ and $128\times 128\times 128$ nm$^3$ simulation cells. Although non-linear, we find a similar trend of $H_c$ with inter-grain exchange coupling and grain sizes. For a fixed value of grain size, qualitatively, $H_c$ decreases with inter-grain exchange coupling, while it is the opposite with grain size. That inter-grain exchange coupling may not be uniform in actual material due to the void or imperfection between the grains, which affects the perfect spin alignment resulting in $M_s$. A larger inter-grain exchange coupling further reduces $K_1$ and $H_c$. With an increase in grain size,  $H_c$ experiences a decrease, similar to that reported in the Dy-substituted neo-magnet\cite{SasakiScripta016}.

Various combinations of grain size and inter-grain exchange coupling reproduce the actual coercivity. For instance, with grain size $\sim 30$ nm, and a $0.58$ ratio for inter-grain to intra-grain \gls{aex} reproduces the realistic \gls{hc} of 0.40 T, as demonstrated in Fig.~\ref{fig:coercivity_mod}. Strictly speaking, a larger inter-grain coupling tends to reduce the $H_c$, regardless of grain size. This suggest that 
we can explore the other magnetic parameters $A_{ex}$ and grain size by using ML-predicted results employing micromagnetic simulations, which demonstrates the fundamental usefulness of \gls{ml}, not just as a black box, but also as a pre-processing input to a micromagnetic model to reverse-engineer the micromagnetic structure.

\section{Conclusion}
Experimental dataset input and \gls{ml} modeling coupled with \gls{dft} predicts \gls{hc} with far greater accuracy and speed than was previously possible using micromagnetic modeling. This technique provides a robust computational foundation for predicting novel permanent magnet materials and optimizing their properties. Using micromagnetic simulation, we first studied magnetization as a function of the applied magnetic field for real and hypothetical magnetic materials, mainly focusing on  1\,:\,5 and 2\,:\,14\,:\,1 rare-earth-based permanent magnets such as Nd$_2$Fe$_{14}$B. Calculations of \gls{hc} based on the hysteresis loop overestimate by a factor of $\sim 5$; the Brown Paradox. The paradox is side-stepped by the judicious use of \gls{ml}, obtaining a more accurate prediction of the target variable by learning its dependence on the input features. We find that \gls{hc} is directly proportional to \gls{ku}, inversely proportional to \gls{ms}, and weakly correlated with \gls{aex}.

We apply the \gls{ml} modeling to Nd$_2$Fe$_{14}$B by first computing its independent variables with \gls{dft}  calculations. 
These calculations show that the $\mathrm{Nd}$ $4g$-sites mainly contribute to the uniaxial magneto-crystalline anisotropy. They also yield a value for the \gls{ct}, which agrees with the experiment. The \gls{dft} predictions suggest the possibility of tuning rare-earth magnetic properties by substituting non-critical elements at specific sites. For instance, Ce-doping at $4f$ sites shows only a slight reduction in \gls{hc}, consistent with the ML prediction. Finally, we engineer the grain boundary size and inter-exchange coupling with the aid of ML-predicted $H_c$, which indicates that the reduction in inter-grain exchange coupling reduces $H_c$. On the other hand, reducing the grain size increases $H_c$ qualitatively.

\acknowledgements{This work is supported by the Critical Materials Innovation Hub, an Energy Innovation Hub funded by the US Department of Energy, Office of Energy Efficiency and Renewable Energy, Advanced Materials and Manufacturing Technologies Office. The Ames National Laboratory is operated for the U.S. Department of Energy by Iowa State University of Science and Technology under Contract No. DE-AC02-07CH11358. GNN acknowledges support from the U.S. Department of Energy, Office of Science, Office of Workforce Development for Teachers and Scientists, Office of Science Graduate Student Research (SCGSR) program. The SCGSR program is administered by the Oak Ridge Institute for Science and Education for the DOE under contract number DE‐SC0014664.}

\appendix{}

\section{\gls{ml} model and hyperparameter tuning}
In this appendix we provide  information about the hyperparameters used in the fine tuning of the \gls{ml} models. With the experimental data, we used the hyperparameters given in Table~\ref{hyperGB} for tuned \gls{xgb} model training.

\begin{table}
\centering
\caption{Hyperparameters used for \gls{xgb} model training of experimental magnetic materials.}\label{hyperGB}
\begin{ruledtabular}
\begin{tabular}{c}
n\_estimators=500\\
min\_weight\_fraction\_leaf=0\\
max\_depth=5\\
learning\_rate=0.006\\

\end{tabular}
\end{ruledtabular}
\end{table}

\begin{table}[b]
\centering
\caption{Fine-tuned hyperparameters for different \gls{ml} models used in the training of micromagnetic simulated  data.}\label{tab:hypertable}
 \begin{ruledtabular}
\begin{tabular}{c|c}
Model &  Hyperparameters \\
\hline
 Tuned \gls{rf} & n\_estimators=1200, max\_depth=5, \\
 &random\_state=1, \\
 &min\_samples\_leaf=3, bootstrap=True, \\
 &max\_features='auto'\\
 \hline 
Tuned \gls{xgb} & n\_estimators=1200, max\_depth= 5, \\
&min\_weight\_fraction\_leaf=0, \\
& learning\_rate=0.007\\
\hline 
Tuned \gls{xgb} & n\_estimators= 1200,  max\_depth= 5, \\
    &min\_child\_weight=0, learning\_rate=0.007, \\
    & random\_state=42\\
\end{tabular}
\end{ruledtabular}
\end{table}
In \mumax data training we use the  hyperparameters given in Table~\ref{tab:hypertable}.

\begin{figure}
\centering
\includegraphics[scale=0.475]{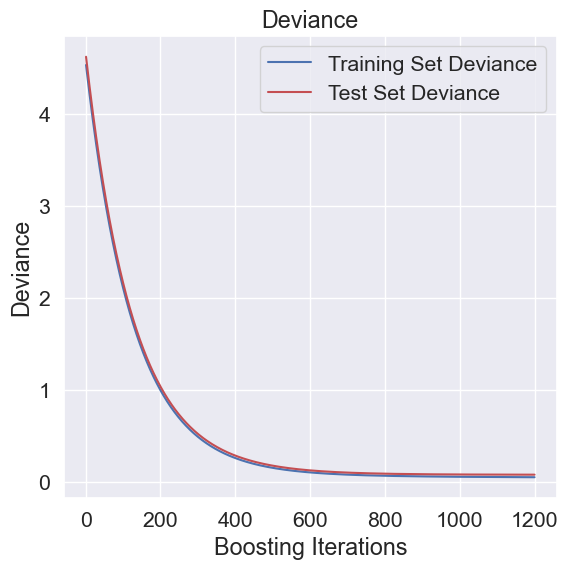}
\caption{Deviance as a function of boosting iterations for the micromagnetic simulated dataset with \gls{xgb} model. Deviance is \gls{mse} between the actual and predicted data set computed in each iteration. Both the training and test data set converge well  after 600 boosting iterations.}
\label{fig:deviance}
\end{figure}

The hyperparameter used in the fine-tuned GBR model training is shown in Fig.~\ref{fig:deviance}. The deviance (\gls{mse}) converges well in both the test and training datasets after 600 boosting iterations, confirming a very good model performance.

\section{DFT parameters for $1:5$ and $1:12$ intermetallics}
Table \ref{DFTtable} provides the DFT computed parameters for $1:5$ and $1:12$ intermetallic compounds.
\begin{table}
\caption{DFT computed intrinsic magnetic parameters for $1:12$ compositions taken from our previous work \cite{BhandariPRR022} except for $1:5$ type materials as referred in Table. $A_{ex}$ is kept fixed for all materials.}\label{DFTtable}
\begin{ruledtabular}
\begin{tabular}{l|ccc}
Material & $M_s$(A/m) & $A_{ex}$(pJ/m) & $K_u$(MJ/m$^3$)\\
\hline
SmCo$_5$\cite{LarsonPRB03}	&1069768.213	&7	&40.27\\
YCo$_5$\cite{SakuraiPRM018}	& 770449.2281	&7	&2.65\\
CeZrFe$_{11}$	& 597220.2602	&7	&7.36\\
CeZrFe$_{11}$N	&632804.6376	&7	&3.42\\
CeTiFe$_{11}$	&587261.7154	&7	&3.29\\
CeTiFe$_{11}$N	&625792.1247	&7	&3.20\\	
\end{tabular}
\end{ruledtabular}
\end{table}

\bibliography{library, mumax}

\end{document}